\documentclass[final,twocolumn,11pt]{elsarticle}

\usepackage[utf8x]{inputenc}
\usepackage{graphicx}
\usepackage{amssymb}
\usepackage{lineno}
\usepackage{textcomp}
\usepackage{enumerate}
\usepackage{anysize}
\usepackage{vmargin}
\usepackage{setspace}
\usepackage{hhline}
\usepackage[usenames]{color}
\usepackage{dcolumn}
\usepackage{bm}
\usepackage{soul}
\usepackage{float}

\begin{document} 

\title{Pure phase BiFeO$_3$ thin films sputtered over Si: A new route towards high magnetization}


\author[cbpf]{G.A. Gomez-Iriarte}
\ead{grecia@cbpf.br}
\author[cbpf]{C. Labre}
\author[nano]{L. A. S. de Oliveira}
\author[cbpf]{J. P. Sinnecker}

\address[cbpf]{Centro Brasileiro de Pesquisas F\'{i}sicas, Rua Xavier Sigaud 150, 22290-180 Rio de Janeiro, RJ, Brazil}

\address[nano]{N\'ucleo Multidisciplinar de Pesquisas em Nanotecnologia - NUMPEX-NANO, Universidade Federal do Rio de Janeiro, Est. de Xer\'em 27, 25245-390, Duque de Caxias, RJ, Brazil}

\date{\today}


\begin{abstract}
We have investigated the structural and magnetic properties of BiFeO$_3$ (BFO) thin films grown over (100)-oriented Si substrates by rf magnetron sputtering in a new route under O$_2$ free low pressure Ar atmosphere. Single-phase BFO films were deposited in a heated substrate and post-annealed in situ. The new routed allows high deposition rate and produce  polycrystalline BFO pure phase, confirmed by high resolution X-ray diffraction. Scanning electron and atomic force microscopy reveal very low surface roughness and mean particle size of 33 nm. The BFO phase and composition were confirmed by transmission electron microscopy and line scanning energy-dispersive X-ray spectroscopy in transmission electron microscopy mode. The surface chemistry of the thin film, analyzed by X-ray photoelectron spectroscopy, reveals the presence of Fe$^{3+}$ and Fe$^{2+}$ in a 2:1 ratio, a strong indication that the film contains oxygen vacancies. An hysteretic ferromagnetic behavior with room temperature high saturation magnetization $\sim 165 \times 10^3$ A/m was measured along the film perpendicular and parallel directions. Such high magnetization, deriving from this new route, is explained in the scope of oxygen vacancies, the break of the antiferromagnetic cycloidal order and the increase of spin canting by change in the surface/volume ratio. Understanding the magnetic behavior of a multiferroic thin films is a key for the development of heterogeneous layered structures and multilayered devices and the production of multiferroic materials over Si substrates opens new possibilities in the development of materials that can be directly integrated into the existent semiconductor and spintronic technologies.
\end{abstract}


\maketitle

\section*{1. Introduction}

Multiferroic materials combines at less two ferroics properties, either coupled or not, being those most interesting the ones that combine magnetic order with electric order. The control of magnetic properties in non multiferroic materials is usually a high energy consumption process involving applied magnetic fields. One the other hand, magnetic control by means of electric fields in multiferroics consumes less energy\cite{Scott2007,Fiebig2016,Spaldin2017} and opens new possibilities in the development of materials that can be directly integrated into the existent semiconductor and spintronic technologies\cite{Scott2007,Fiebig2016,Spaldin2017,Spaldin2005,Bea2008}. 

The BiFeO$_{3}$ (BFO) compound is one of the most studied multiferroics material, either in bulk, nanoparticles or in thin films\cite{Ramesh2007}. BFO bulk materials have a g-type antiferromagnetic structure with a cycloid spin structure period of approximately 62 nm along (110) direction\cite{kuo2016single}\cite{ramazanoglu2011}. The synthesis process of pure BFO is usually difficult due the bismuth volatility and the appearance of spurious phases like Bi$_{2}$Fe$_{4}$O$_{9}$,  Bi$_{25}$FeO$_{39}$, Bi$_{2}$O$_{3}$ and Fe$_{2}$O$_{3}$. Thin film fabrication of pure BFO by pulsed laser deposition (PLD) or by rf magnetron sputtering normally involves the use of an oxygen rich atmosphere\cite{Mori2017} to reduce the formation of oxygen vacancies that are reported to degrade some important ferroelectric properties\cite{Yang2010}. Nevertheless this may also increase the presence of the undesirable spurious phases\cite{Mori2017}. In this work, high magnetic pure polycrystalline BFO thin films were grown over (100)-oriented Si substrates using a new route, with rf magnetron sputtering using a O$_2$ free atmosphere. The surface chemistry of the thin film, analyzed by X-ray photoelectron spectroscopy, reveals the presence of Fe$^{3+}$ and Fe$^{2+}$ in a 2:1 ratio which, in the absence of spurious Fe oxides like Fe$_3$O$_4$, is a strong indication that the film contains oxygen vacancies. An hysteretic ferromagnetic behavior with room temperature high saturation magnetization $\sim 165 \times 10^3$ A/m is measured along the film perpendicular and parallel directions. Such high magnetization is explained in the scope of oxygen vacancies as well as the break of the antiferromagnetic cycloidal order and the increase of spin canting by change in the surface/volume ratio due to the crystallite size of the order of 30 nm. Understanding the magnetic behavior of a multiferroic thin films is a key for the development of heterogeneous layered structures and multilayered devices, e.g. multiferroic tunnel junctions and for multiferroic exchange bias heterostructures, e.g. in Magnetoelectric Random Acess Memories (MERAMS). The production of multiferroic materials over Si substrates opens new possibilities in the development of materials that can be directly integrated into the existent semiconductor and spintronic technologies.

\section*{2. Experimental}
\subsection*{2.1 New sputtering route} 

BFO thin films were grown directly over (100)-oriented Si substrates by rf magnetron sputtering, with a base pressure of 1.3$\times 10^{-6}$ Pa (10$^{-8}$ Torr), using a BFO commercial target (AJA International, Inc.). All substrates were previously cleaned with hydrofluoric acid in order to remove the thermal SiO$_{2}$ layer and induce an hydrophobic surface. Deposition of the BFO film was made using 35 W of rf power at a pure Ar atmosphere working pressure of 4$\times 10^{-1}$ Pa (3$\times 10^{-3}$ Torr). Usually BFO is deposited under O$_2$ atmosphere \cite{Khalkhali2014,ranjith2009}. In this new route, the atmosphere is O$_2$ free. The substrate temperature was maintained at 873 K during deposition in order to produce an homogeneous film. A further in situ annealing was made at 973 K for 3600 s, at the base pressure, to induce pure BFO phase formation \cite{Palai2008}.

\subsection*{2.2 Graze incident high resolution X-ray diffraction}

\begin{figure*}[ht]
\centering
\includegraphics[width=0.7\textwidth]{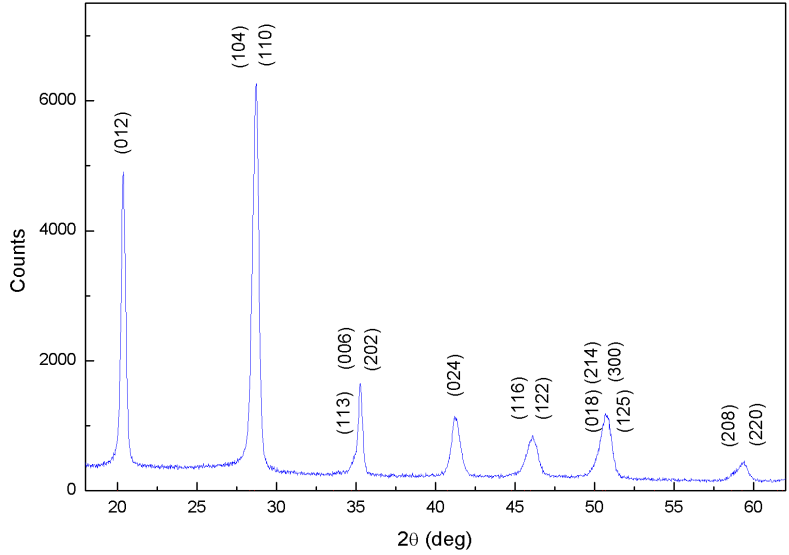}
\caption{High resolution graze incidence X-ray diffraction of pure phase BFO thin film sputtered over (100)-oriented Si substrate. \label{LNLS_GG57A3}}
\end{figure*}

The BFO phase and grain size were characterized by graze incident high resolution X-ray diffraction (XRD) at the LNLS synchrotron in Campinas, SP, Brazil, with 9 keV energy, as shown in Figure \ref{LNLS_GG57A3}.

The XRD diffractogram shows a R3cH polycrystalline BiFeO$_{3}$ pure phase, indexed with JCD188396 card, without any traces of the common spurious phases Bi$_{25}$FeO$_{39}$, Bi$_{2}$Fe$_{4}$O$_{9}$ and iron oxides \cite{Ramirez2015,Chen2009}. No significant amorphous phase signature can be detected in the diffractogram as expected considering the Fe-O and Bi$_2$O$_3$-Fe$_2$O$_3$ phase diagrams \cite{Palai2008,Lad1988,Lanier2009}

The estimated particle size was calculated from Debye-Scherrer equation, considering the peaks overlap and the instrumental contributions, giving a mean size of $\sim 30 \pm 8$ nm (Table \ref{tab:size}). 

\subsection*{2.3 Surface morphology}

\begin{figure*}[ht]
\centering
\includegraphics[width=\textwidth]{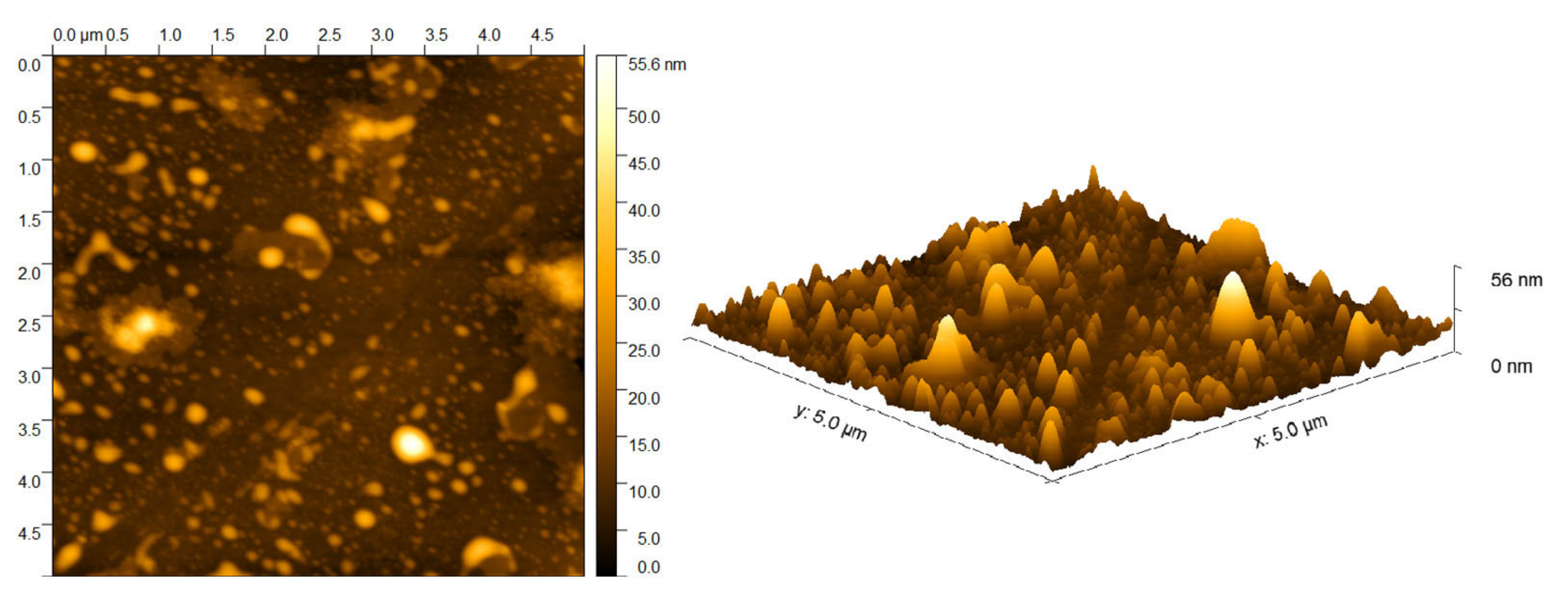}
\caption{AFM of the BFO thin film, showing a low roughness surface.\label{AFMGG57A3}}
\end{figure*}

Surface morphology and grain size were also studied by atomic force microscopy (AFM - Brucker Nonoscope V - LabSurf/CBPF), measured in tapping mode, and by field emission gun scanning electron microscopy (FEG SEM - Raith eLine - LABNANO/CBPF). Atomic force microscopy, shown in Figure \ref{AFMGG57A3}, confirms a low surface roughness average (Ra) around $\sim 4 \pm 1$ nm. 

\begin{figure*}[ht]
\centering
\includegraphics[width=0.6\textwidth]{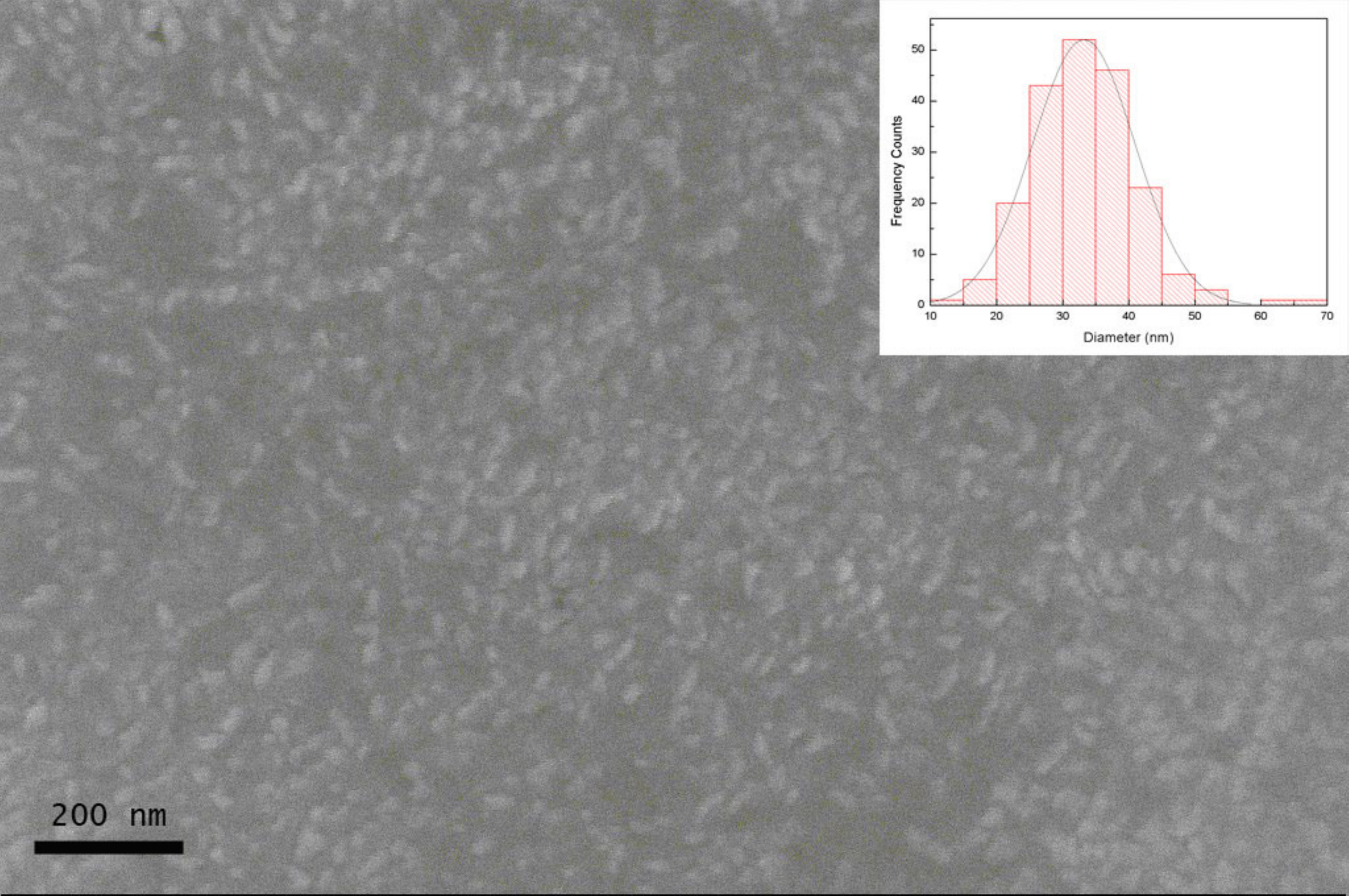}
\caption{SEM image the BFO thin film surface. The inset shows the grain size distribution.\label{SEMGG57A3}}
\end{figure*}

Figure \ref{SEMGG57A3} shows a SEM image of the BFO film confirming a high homogeneous film surface. Figure \ref{SEMGG57A3} inset shows the grain size distribution with a mean size $\sim 33 \pm 7$ nm. The grain size calculated by Debye-Scherrer equation is close to the one found by SEM images as can be seen in Table \ref{tab:size}. Both grain size values are smaller than the antiferromagnetic cycloid ordering ($\sim 60$ nm) in bulk BFO. It is well known that the breaking of the antiferromagnetic cycloid ordering gives rise to a small ferromagnetic contribution to BFO magnetic behavior \cite{kuo2016single}.

\begin{table}[ht]
\centering
\caption{\label{tab:size} Grain size of BFO thin film calculated by Debye-Scherrer equation and obtained by SEM histogram.}
\resizebox{\columnwidth}{!}{
\begin{tabular}{lc}
Method & Grain Size (nm)\\
\hline
Debye-Scherrer equation (XRD) & 30$\pm$8 \\
SEM & 33$\pm$7 \\
\end{tabular}
}
\end{table}

\subsection*{2.4 Transmission electron microscopy}

\begin{figure*}[ht]
\centering
\includegraphics[width=0.7\textwidth]{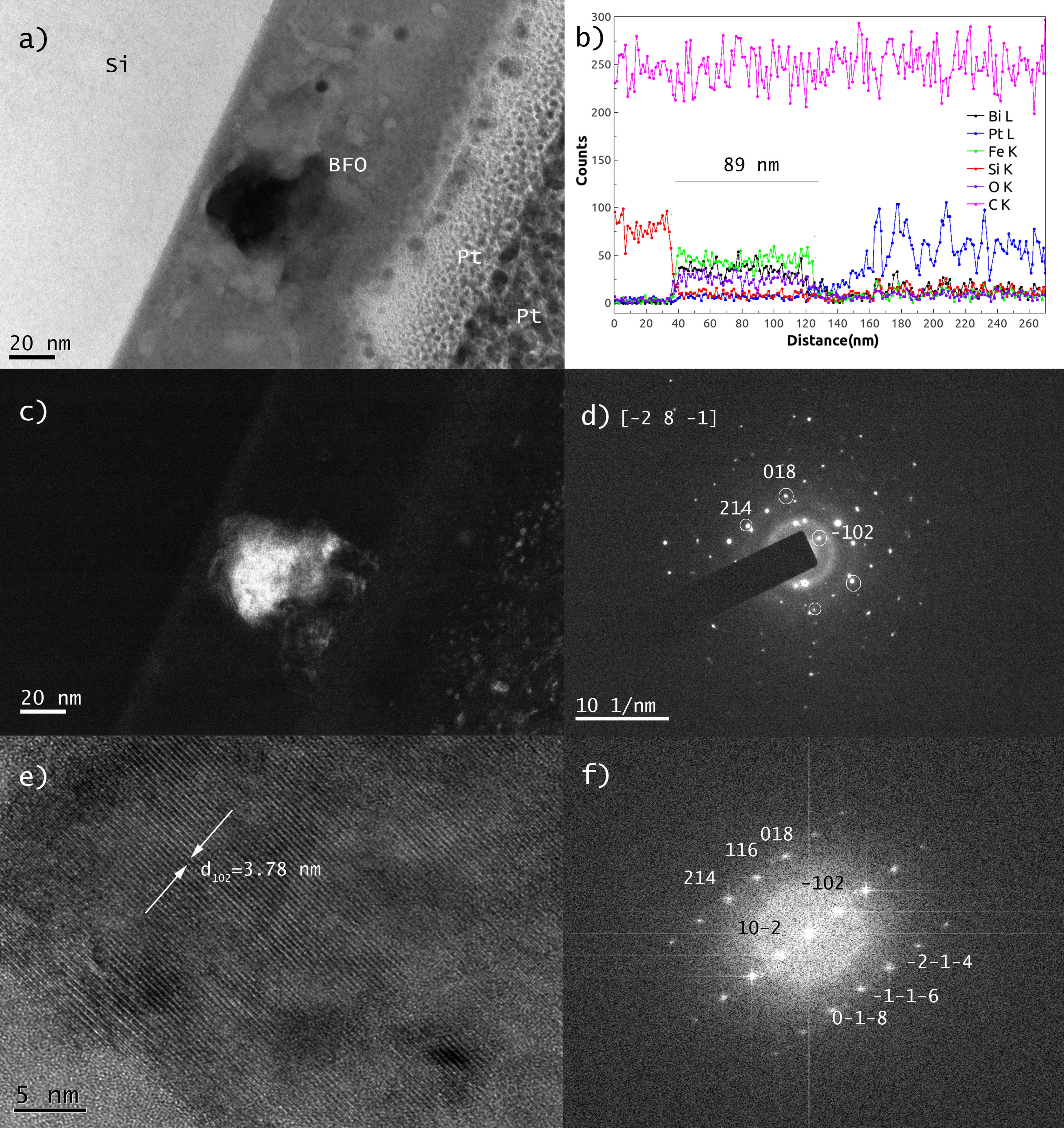}
\caption{a) Bright field TEM image of BFO thin film cross section, b) EDS line scan in STEM mode of the BFO thin film cross section, c) dark field image from (-102) plane, d) SAED obtained in the BFO [-2 8 -1] zone axis, e) HRTEM image from BFO crystallite in the zone axis [-2 8 -1], f) FFT pattern of the HRTEM BFO crystallite in the zone axis [-2 8 -1].\label{GG57A3_STEM_E_DIFF_HRTEM}}
\end{figure*}

Transmission electron microscopy (TEM - Jeol JEM 2100-F - LABNANO/CBPF) was employed to verify the BFO film thickness, film/substrate interface quality and composition. The TEM sample was prepared by focus ion beam (FIB - Tescan Lyra 3 - LABNANO/CBPF), with gallium ion milling, initially with a accelerating voltage of 30 kV and a beam current of 4.97 nA. Eventually those values were changed in few polishing stages (30 kV - 1.2 nA, 30 kV - 0.2 nA, 10 kV - 0.145 nA and 5 kV - 0.02 nA). To reduce Ga implantation during the FIB milling process, thin carbon followed by a thin Pt layers were deposited over the BFO film before the milling. Figure \ref{GG57A3_STEM_E_DIFF_HRTEM}a shows the bright field image of the BFO film cross section where the Si, BFO and Pt rich regions are indicated. As can be observed, the film/substrate interface is very smooth presenting a good adhesion. The BFO film has a total thickness around 89 nm, as also confirmed by electron dispersive spectroscopy (EDS) line scan in scanning transmission electron microscopy (STEM) mode of the BFO thin film cross section. One can observe that up to 23.3 nm from the surface the BFO film presents low crystallinity caused by the milling process. This effect is well known and widely reported in the literature\cite{Siemons2014,RUBANOV20032238}. Figure \ref{GG57A3_STEM_E_DIFF_HRTEM}b presents the EDS line scan in STEM mode showing clearly the substrate/film transition, the $\sim 89$ nm sputtered film composed by Bi,Fe and O, and the film surface transition to the Pt rich region. Carbon was found along the whole scan as expected. The dark field image of the same selected area is presented in Figure \ref{GG57A3_STEM_E_DIFF_HRTEM}c where the bright zones are BFO crystals diffracting. One can see also some bright spots at the Pt rich region. This bright BFO region corresponds to the (-102) diffraction spot on Figure \ref{GG57A3_STEM_E_DIFF_HRTEM}d where the selected area electron diffraction (SAED) is presented. The SAED, obtained in the BFO zone axis [-2 8 -1], shows the BFO spots together with Si and amorphous Pt. High resolution TEM image of one BFO crystallite is observed in Figure \ref{GG57A3_STEM_E_DIFF_HRTEM}e. Atomic planes with inter-planar distances of $\sim 3.78$ nm where indexed as corresponding to the (102) plane. A Fast Fourier Transform (FFT) of the BFO crystallite in the zone axis [-2 8 -1] in Figure \ref{GG57A3_STEM_E_DIFF_HRTEM}e, shown in Figure \ref{GG57A3_STEM_E_DIFF_HRTEM}f, matches the SAED pattern, with the brighter spots corresponding to the {102} planes.

\subsection*{2.5 X-Ray photoelectron spectroscopy}

\begin{figure}[ht]
\centering
\includegraphics[width=0.5\textwidth]{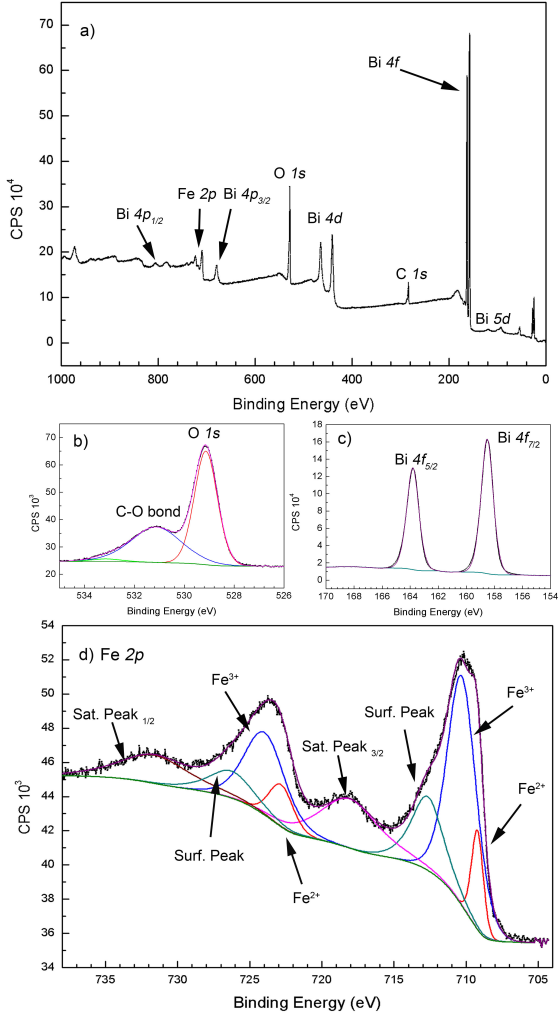}
\caption{XPS spectra for a) survey, b)Oxygen, c)Bi $4f$ and d) Fe $2p$ of the BFO thin film grown in a O$_2$ free low pressure Ar atmosphere .\label{XPSGG57A3}}
\end{figure}

X-Ray photoelectron spectroscopy (XPS) measurements were made using a SPECS PHOIBOS 100/150 spectrometer with a polychromatic Al K$\alpha$ 1486.6 eV radiation and a 150 mm hemispherical analyzer, in order to study the Fe and Bi chemical state at the thin film surface. The emitted photoelectrons were detected using a 0.05 eV energy step for high resolution spectras and 0.5 eV for the survey spectrum. The binding energy scale of the XPS spectra was calibrated using the C $1s$ (284.6 eV). From the XPS survey spectra (Figure \ref{XPSGG57A3}a) the calculated Bi:Fe ratio, considering Fe $2p$ and Bi $4p_{3/2}$ peaks due to a close binding energies and similar intensities, is $\sim 1$, compatible with the BFO stoichiometry. Figures \ref{XPSGG57A3}b,c and d exhibit the high resolution spectra of O $1s$, Bi $4f$ and Fe $2p$ respectively, which were deconvoluted using CASA-XPS software considering a Shirley background. Figure \ref{XPSGG57A3}b shows O $1s$ peak (529.16 eV) indicating a binding energy present when an O-Fe bond exists\cite{poulin2010,zaki2014surface} and a C-O bond peak (531.15 eV), typical for surfaces with any oxygen component. Figure \ref{XPSGG57A3}c shows two Bi $4f$ peaks with binding energies of 158.51 eV and 163.82 eV and separation of 5.31 eV, very close to previous reported values for Bi$^{3+}$ oxidation state\cite{ Marchand2016}. In Figure \ref{XPSGG57A3}d, Fe $2p$ spectra exhibits the Fe $2p_{3/2}$, Fe $2p_{1/2}$ and their respective satellites peaks. The separation between Fe $2p_{3/2}$ (710.39 eV) and Fe $2p_{1/2}$ (724.00 eV) is 13.61 eV and the separation between Fe $2p_{3/2}$ peak and its satellite is 8.08 eV. These results are in accordance with previous reported values for BiFeO$_{3}$ where a Fe$^{3+}$ predominance is observed\cite{Marchand2016,xu2015effect}. The deconvolution of Fe $2p_{3/2}$ cannot be done with a simple gauss-lorentzian fitting because of its asymmetry. A good fit of the asymmetric Fe $2p_{3/2}$ peak (Figure \ref{XPSGG57A3}d) can be achieved only by taking into account a surface peak\cite{Droubay2001} at 712,6 eV, which is consider as Fe$^{2+}$ satellite due to the loss of the crystal field at surface \citep{poulin2010,bhargava2007,wilson2014,grosvenor2004}, in addition to the Fe$^{2+}$ and Fe$^{3+}$ peaks at 709,20 eV \cite{xu2015effect,wilson2014} and 710,31 eV \cite{wilson2014}, respectively.  

Table \ref{tab:xps} summarizes the deconvoluted XPS data. The Fe contribution shows a Fe$^{3+}$:Fe$^{2+}$ ratio of about 2:1 at the thin film surface. Although this ratio is the same found in magnetite phase (Fe$^{2+}$Fe$_2^{3+}$O$_4$), which is a possible magnetic spurious phase in BiFeO$_3$, this is not the case here, as confirmed by the hight resolution X-ray analysis. Therefore, we address the significant presence of Fe$^{2+}$ as the presence of oxygen vacancies that can be a result of the O$_2$ free low pressure Ar atmosphere used in our sample preparation.
\begin{table}[ht]
\centering
\caption{\label{tab:xps} Fe $2p$ peaks deconvolution.}
\resizebox{\columnwidth}{!}{
\begin{tabular}{l c c c }
& \makebox[3cm][c]{Binding energy (eV)} & \makebox[2.5cm][c]{Composition}& \makebox[2cm][c]{Reference}\\
\hline
Fe $2p_{3/2}$ & 709.21 & Fe$^{2+}$ &\cite{xu2015effect},\cite{wilson2014}\\
			  & 710.31 & Fe$^{3+}$&\cite{wilson2014}\\
              & 712.68 & Surface Peak&\cite{poulin2010}\\
              & 718.08 & Satellite Peak&\cite{wilson2014}\\
Fe $2p_{1/2}$ & 722.81 & Fe$^{2+}$&\\
        	  & 723.94 & Fe$^{3+}$& \\
       		  & 726.28 & Surface Peak&\\
              & 731.63 & Satellite Peak&\\
\end{tabular}
}
\end{table}

\subsection*{2.6 Magnetic Properties}

Magnetization was measured as a function of applied magnetic field, for different temperatures, using a Quantum Design MP3M SQUID in VSM mode. The magnetic hysteresis curves (MH) were obtained with a maximum applied magnetic field of 7 T and temperatures ranging from 2 to 300 K. Measurements were performed with either an in-plane or out-of-plane applied magnetic field, and the diamagnetic contribution from the Si substrate was removed from the data. All measurements were corrected for the residual remanent field of the SQUID superconducting magnet using a paramagnetic Pd standard sample, to avoid experimental artifacts.

\begin{figure}[H]
\centering
\includegraphics[width=0.5\textwidth]{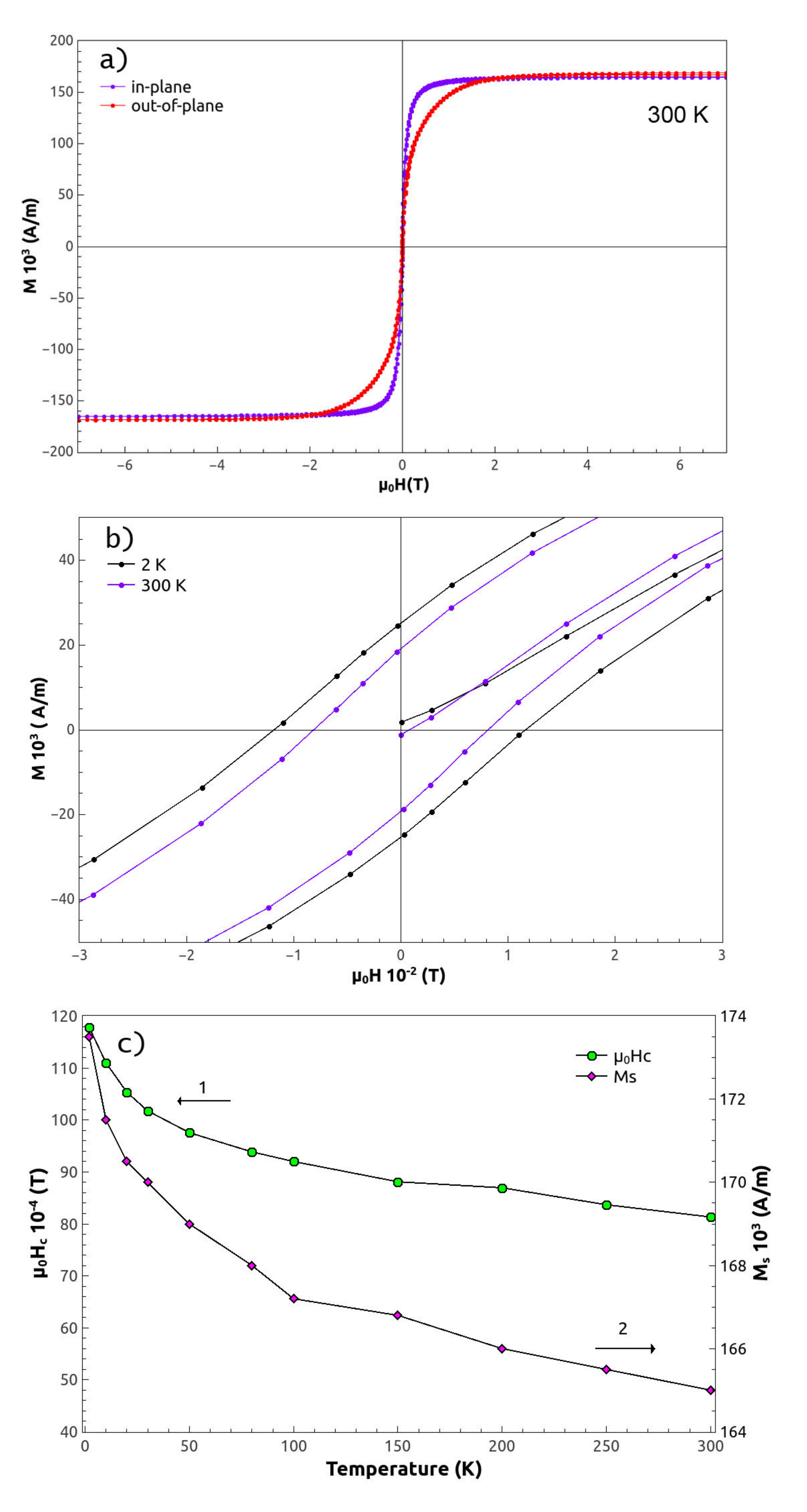}
\caption{a) Room temperature in-plane and out-of-plane magnetic hysteresis curves of pure BFO thin film over (100)-oriented Si, b) detail of the in-plane magnetic hysteresis curves measured at 2 and 300 K, c) coercive field (left axis) and saturation magnetization (right axis) behavior as function of temperature.\label{Magneticas_GG57A3}}
\end{figure}

Figure \ref{Magneticas_GG57A3}a shows the room temperature magnetic hysteresis curves measured in-plane and out-of plane. As expected, the out-of-plane curve present a shape anisotropy contribution due to the demagnetizing field. Both in-plane and out-of-plane present room temperature high saturation magnetization of $\sim 165\times 10^3$ A/m, corresponding to $\sim 1 \mu_B$ per unit cell. Wang {\it et al.} found a saturation magnetization value of $150 \times 10^3$  A/m for a 70 nm BFO thin film grown epitaxially on SrTiO$_3$ \cite{Wang2003}. A saturation magnetization of $180  \times 10^3$A/m for a 40 nm thick was also reported by Laughlin {\it et al.} on BFO thin films grown on SrTiO$_3$/Si by MBE \cite{laughlin2013magnetic}. Those values are much higher than the 28 A/m (38 nm thick) epitaxially grown over LaAlO$_3$ film reported by Cheng {\it et al.} \cite{Cheng2011}. It is important to stress that all those high magnetization reported values were obtained for epitaxially grown thin films, contrary to our polycrystalline sample. As can be seen in detail at Figure \ref{Magneticas_GG57A3}b, a non-null in-plane magnetization at remanence ($\sim 20 \times10^3$ A/m) and coercive field ($\sim 86 \times10^{-4}$ T) was found even at room temperature. The behavior of the in-plane saturation magnetization and coercive field as function of temperature can be seen on Figure \ref{Magneticas_GG57A3}c. Both behaviors does not change significantly with increasing temperature, as also observed by Wang {\it et al.} \cite{Wang2003}. 

\section*{3. Discussion}

In literature, one can find many reports about weak ferromagnetic behavior in pure BFO nanostructured materials (nanoparticles \cite{park2007size}, nanowires \cite{deOliveira2012}, nanotubes \cite{DEOLIVEIRA20131593} and thin films \cite{Wang2003,Eerenstein2005,Cheng2011}). However, this topic is still in debate and some authors attribute this weak ferromagnetism to the presence of spurious magnetic phases such as Bi$_{25}$FeO$_{39}$, Bi$_{2}$Fe$_{4}$O$_{9}$ and $\gamma-$Fe$_2$O$_3$   \cite{Ramirez2015,Vijayanand2009,Mori2017}. Based on our characterizations, there are no spurious magnetic phases on the rf sputtered BFO over Si samples. Weak ferromagnetism on pure BFO is commonly attributed either to the suppression of the spin cycloidal order, to the increase of the spin canting by change in the surface/volume ratio or to the presence of oxygen vacancies\cite{laughlin2013magnetic,DEOLIVEIRA20131593,deOliveira2012}. Although our film is $\sim$ 89 nm thick, it is clearly composed by crystallites of 30 nm mean size, i.e. bellow the 62 nm long cycloidal spin structure, giving rise to uncompensated spins \cite{park2007size}. Besides that, the O$_2$ free low Ar pressure atmosphere promotes oxygen deficiency, leading to Fe$^{2+}$ state and could give rise to carrier-mediated local ferromagnetic order across Fe$^{3+}$-O$^{2-}$-Fe$^{2+}$.

\section*{4. Summary}

In summary, we have grown polycrystalline BFO thin films by rf sputtering in a O$_2$ free low pressure Ar atmosphere over (100)-oriented Si substrates. The pure BFO is confirmed by high resolution X-ray diffraction which shows no trace of commonly observed spurious magnetic phases. SEM and AFM reveals a very low roughness surface (less than 4 nm) and mean particle size of 30 nm. The BFO phase and composition were confirmed by TEM and XPS, the latter also confirming the presence of Fe$^{2+}$. Hysteretic ferromagnetic behavior with high saturation magnetization of $\sim 165\times 10^3$ A/m and non-zero coercivity at room temperature was measured along the film perpendicular and parallel directions. Such high magnetization has been explained in the scope of the suppression of the spin cycloidal order, increase of the spin canting and the presence oxygen vacancies. The fabrication of high magnetic pure BFO over Si substrates is an important step towards the integration of magnetic multiferroic oxides with semiconductors. This is a key for the development of heterogeneous layered structures and multilayer devices with high impact on technological advancements, e.g. magnetoelectric random access memories, and for multiferroic materials direct integration into the existent semiconductor and spintronic technologies.
 
\section*{Acknowledgments}

The authors thank Dr. Arbelio Pentón-Madrigal for the helpful discussion of the X-ray diffraction data. The authors acknowledge the support of Conselho Nacional de Desenvolvimento Científico e Tecnológico - CNPq under Grants No.484800/2013-2 and No.305667/2014-9, FAPERJ under Grant PENSARIO No. E-26/010.002996/2014 and Coordenação de Aperfeiçoamento de Pessoal de Nível Superior CAPES. The authors acknowledge the use of the Brazilian National Laboratory of Nanoscience and Nanotechnology (LABNANO/CBPF) and LabSurf/CBPF for the sample preparation, TEM, SEM, AFM and XPS characterizations, and LNLS/CNPEM Campinas for the high resolution X-rays under proposal 20160352.  

\section*{References}

\bibliography{bfoMMM}
\bibliographystyle{unsrt}

\end{document}